\documentclass[preprint,11pt]{elsarticle}
\usepackage{graphicx}
\usepackage{amssymb}
\usepackage{lineno}
\usepackage{fancyhdr}
\usepackage{epsfig}
\usepackage{epic}
\usepackage{eepic}
\usepackage{amsmath}
\usepackage{mathtools}
\usepackage{threeparttable}
\usepackage{amscd}
\usepackage{here}
\usepackage{graphicx}
\usepackage{subfigure}
\usepackage{amsmath}
\usepackage{graphicx}
\usepackage{dcolumn}
\usepackage{bm}
\usepackage{amssymb}
\usepackage{latexsym}
\usepackage{bm}
\usepackage{multirow, array}
\usepackage[raggedright]{sidecap}
\usepackage{float}
\usepackage{lscape}
\usepackage{tabularx}
\usepackage[export]{adjustbox}
\usepackage{longtable}
\usepackage{colortbl}
\usepackage{rotating}
\usepackage{comment}
\usepackage{natbib}
\usepackage{appendix}
\begin{document}
\begin{frontmatter}




\title{Application of Flexible Presentation of Quantum Images in Multipartite Correlations}

\author{Mario Alberto Mercado Sanchez$^{1}$, Guo-Hua Sun$^{2}$ and Shi-Hai Dong$^{1}$}
\address{$^{1}$Laboratorio de Informaci\'{o}n Cu\'{a}ntica, CIDETEC, Instituto Polit\'{e}cnico Nacional, UPALM, CDMX 07700, Mexico}
\ead{ometitlan@gmail.com (M. A. Mercado-Sanchez), sunghdb@yahoo.com (G. H. Sun), dongsh2@yahoo.com (Corresponding author: dongsh2@yahoo.com. Tel: +52-55-57296000 ext. 52522.)}
\address{$^{2}$Catedr\'{a}tica CONACYT, CIC, Instituto Polit\'{e}cnico Nacional, CDMX 07700, Mexico}

\begin{abstract}
We apply quantum model inspired on the classical Bayesian method also called mutual information to study the multipartite correlation in quantum images by using the flexible representation of quantum images (FRQI). This can be reflected by considering von Neumann entropy. The results are compared between two images of size $2\times 2$ and $8\times 8$ from different classical and quantum methods. We find that the classical joint entropy is invariant under transformation of change of color but the quantum entropy is sensitive to this change. It is shown that the total correlation $I_T$  could arrive to the double amount of the classical joint entropy.

\end{abstract}


\end{frontmatter}

\section{Introduction}
Quantum computation has become an important and effective tool to overcome the high computational requirements of classical digital image processing in terms of the characteristics of parallel computing offered by the phenomenon of quantum superposition, which is an important fundamental principle in quantum mechanics. The entanglement is another important characteristic in the quantum correlations \cite{phoenix_quantum_2015}. These two quantum characteristics stimulated  Feynman \cite{feynman_simulating_1982} to propose the quantum computing model. The advance on this new science has provided us for two important algorithms, i.e., the Shor algorithm \cite{shor_algorithms_1994}, which is used to factorize integer number in polynomial in a non-exponential time and the Grover database search algorithm \cite{grover_fast_1996}, which can be used to explore amount of data. Since then quantum computing has been generalized to other areas such as information theory, cryptography, mechatronics, image processing, chemistry, neural networks and others \cite{daoyi_dong_robust_2012, gupta_quantum_2017, benioff_quantum_1998, carleo_solving_2017, chen_adaptive_2014, ulyanov_cognitive_2017}. In this work, however, we focus on the quantum images \cite{yan_survey_2016} which began in 1997 when Vlaso proposed a method called orthogonal image algorithm to find specific patterns in binary images \cite{Vla}. After that, Venegas-Andraca proposed a quantum image model named \textit {Qubit Lattice Representation} to encode quantum images \cite{venegas-andraca_storing_2003}. Latorre {\it et al.} \cite {Lat} proposed a model called \textit {Real Ket Representation}, which has a special interest in the compression of information, i.e., the images are those quantum states which take levels of gray as the coefficients of the states.

The \textit {Flexible Representation for Quantum Images} (FRQI) model proposed in 2010 \cite{le_flexible_2011} is a new approach to represent the color and the corresponding position for an image in a normalized wave function. It should be recognized that this model represents a significant reduction in the number of qubits needed to encode the image, that is, this model only requires $2n+1$ qubits \cite{le_flexible_2011} and is unlike the \textit {Qubit Lattice Representation}, which requires $n^2$ qubits to store an image of $n\times n$ pixels. Up to now, there are wide applications of this model to quantum films \cite{iliyasu_framework_2011}, authentication of images by watermark \cite{iliyasu_watermarking_2012} or data mining \cite{yan_quantum_2012}. In 2013 Yan proposed a model for comparison of multiple pairs of images \cite{yan_parallel_2013}. Moreover, other relevant contributions are also made to multipartite systems via quantum mutual information \cite{phoenix_quantum_2015, sazim_quantum_2016, Lau}.
In this work, we are about to study the entanglement phenomena in a quantum codification of two images in two dimensions by means of entanglement measure such as entropy or mutual information\footnote{These two measures are also used in classical image fusion that is an important application in medical physics. \cite{gupta_mutual_2008, Cal, hill_medical_2001}.} used in the context of quantum image processing, \textit {Quantum Information Processing} (QuIP) and \textit {Quantum Signal Processing} (QuSP) \cite{Eld}. We focus on encoding the information of a pair of 2D images of size $2\times2$ and $8\times8$ in terms of the FRQI model and analyze the behavior of the joint entropy and mutual information in the proposed function to verify the existing correlations between the images.

The plan of this work  is organized as follows. In Sec. II we give a brief review of FRQI and then apply this method to an image. We will study the process of comparing two images in Sec. III. Two typical examples such as the comparison of two images of size $2\times2$ and $8\times 8$ are illustrated in Sec. IV. Some concluding remarks are summarized in Sec. V.

\section{FRQI and its application to an image}
The FRQI model was proposed for a normalized quantum state \cite{le_flexible_2011}. The information about the color and the position of each pixel is encoded in the following way:
\begin{equation}\label{1}
\begin{array}{l}
|\psi\rangle =\displaystyle\frac{1}{2^{n}} \sum\limits_{i=0}^{2^{2n}-1} |c_{i}\rangle\otimes |i \rangle\\[3mm]
~~~~=\displaystyle\frac{1}{2^{n}} \sum\limits_{i=0}^{2^{2n}-1} (\cos \theta_{i} |0 \rangle + \sin \theta_{i} |1 \rangle) \otimes |i \rangle,
\end{array}
\end{equation}
where, $\theta=(\theta_{1}, \theta_{2}, ..., \theta_{2^{2n}-1})$,  with $\theta_{i}\in [0, \pi/2]$, is the angle vector that encodes the color information. It should be pointed out that the notation $|c_{i}\rangle=(\cos \theta_{i} |0 \rangle + \sin \theta_{i} |1 \rangle)$ denotes the color of each pixel in an image and the set of states $|i\rangle =|0\rangle , |1\rangle , |2\rangle, ..., |2^{2n}-1\rangle$ represent the position in a sequence of base states of a number of qubits. In the calculation, we always use the basic computation bases $|0\rangle=\left(\begin{array}{l}1\\0\end{array}\right)$ and $|1\rangle=\left(\begin{array}{l}0\\1\end{array}\right)$, which correspond to a qubit. It is easy to verify that the wave function $|\psi\rangle$ is normalized, i.e.,
\begin{equation}
|| |\psi\rangle ||=\frac{1}{2^{n}} \sqrt{\sum_{i=0}^{2^{2n}-1} (\cos ^2 \theta_{i} + \sin ^2 \theta_{i})}=1.
\end{equation}

\begin{figure}[h]
\centering
\includegraphics[width=5cm]{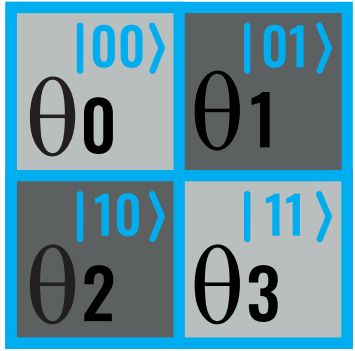}
\caption{ An image of $2\times 2$ pixels with 1 qubit in color and 2 qubits in position.}
\label{Fig:EjemFRQI.png}
\end{figure}

Let  us illustrate how to construct the wave functions with one qubit in color and two qubits in position as illustrated in Fig.\ref{Fig:EjemFRQI.png}. For example, the wave function that encodes the image of four pixels can be expressed as $|\psi\rangle=\frac{1}{2}[(\cos \theta_{0}|0\rangle+\sin \theta_{0}|1\rangle)\otimes |00\rangle+(\cos \theta_{1}|0\rangle+\sin \theta_{1}|1\rangle)\otimes |01\rangle+(\cos \theta_{2}|0\rangle+\sin \theta_{2}|1\rangle)\otimes |10\rangle+(\cos \theta_{3}|0\rangle+\sin \theta_{3}|1\rangle)\otimes |11\rangle]$.

Now, we show how to use FRQI model in a normalized wave function $|\psi\rangle$ and then to obtain the density matrix $ \rho $, from which we are able to calculate the von Neumann entropy.
According to this model, the wave function is composed of the qubit of color $q_{\rm c}$ characterized by the parameter $\theta\in[0, \pi/2]$ and the mesh of qubits of position $| i \rangle$. Using a transformation $[0, 255]$ to the domain $[0, \pi /2]$ we have
$\theta_{i}=\text{(color)}_{i}\times\frac{\pi}{2}\times\frac{1}{255}$ and $q_{c}=|c_{i}\rangle=\sin (\theta_{i} ) \left|1\right\rangle +\cos (\theta_{i} ) \left|0\right\rangle$. Since the images are the size $ 2 \times 2 $ in two dimensions we need a 4 dimensional vector $\vec\theta$. The number of qubits used to encode the position information is two qubits, that is, $|i \rangle=[|0 \rangle, |1 \rangle, |2 \rangle, |3 \rangle]=[|00 \rangle, |01 \rangle, |10 \rangle, |11 \rangle]$. For example, if taking a color configuration as $\{51, 204, 204, 51\}$, then we obtain the color vector $\vec\theta$ and the vector $q_{c}$ that represents the qubit that encodes the color information respectively as
\begin{equation}
\vec\theta = \begin{bmatrix}
 0.314 \\1.256 \\  1.256 \\  0.314
\end{bmatrix}, ~~~~~~~
q_{c} = \begin{bmatrix}
 0.951 \left|0\right\rangle +0.309 \left|1\right\rangle \\0.309 \left|0\right\rangle +0.951 \left|1\right\rangle \\ 0.309 \left|0\right\rangle +0.951 \left|1\right\rangle \\ 0.951 \left|0\right\rangle +0.309 \left|1\right\rangle
\end{bmatrix}.
\end{equation}
The normalized wave function that encodes the color and position information of the image is constructed by considering Eq. (\ref {1})
\begin{equation}
\begin{array}{l}
|\psi\rangle =\displaystyle\frac{1}{2} \sum\limits_{i=0}^{3} (\sin (\theta_{i} ) \left|1\right\rangle +\cos (\theta_{i} ) \left|0\right\rangle) \otimes |i \rangle\\[3mm]
~~~~= 0.475 |000\rangle +0.154 |001\rangle +0.154 |010\rangle +0.475 |011\rangle\\[2mm]
~~~~~~~+0.154 |100\rangle +0.475 |101\rangle +0.475 |110\rangle +0.154 |111\rangle,
\end{array}
\end{equation}
from which we have the density matrix $\rho$
\begin{equation}
\begin{array}{l}
\rho=|\psi \rangle \langle \psi |\\
~~=\left(
\begin{array}{cccccccc}
 0.226 & 0.073 & 0.073 & 0.226 & 0.073 & 0.226 & 0.226 & 0.073 \\
 0.073 & 0.023 & 0.023 & 0.073 & 0.023 & 0.073 & 0.073 & 0.023 \\
 0.073 & 0.023 & 0.023 & 0.073 & 0.023 & 0.073 & 0.073 & 0.023 \\
 0.226 & 0.073 & 0.073 & 0.226 & 0.073 & 0.226 & 0.226 & 0.073 \\
 0.073 & 0.023 & 0.023 & 0.073 & 0.023 & 0.073 & 0.073 & 0.023 \\
 0.226 & 0.073 & 0.073 & 0.226 & 0.073 & 0.226 & 0.226 & 0.073 \\
 0.226 & 0.073 & 0.073 & 0.226 & 0.073 & 0.226 & 0.226 & 0.073 \\
 0.073 & 0.023 & 0.023 & 0.073 & 0.023 & 0.073 & 0.073 & 0.023 \\
\end{array}
\right)
\end{array}
\end{equation}
from which we have $\text {Tr} (\rho ^ 2) = 1 $. Thus, the von Neumann entropy is zero.

Since $\rho$ represents a pure state composed of three qubits, i.e., one qubit of color and two qubits that encode the position, $\rho = \rho_{\rm color, position} $ a bipartite system.
The entropies of each subsystem for a pure system are equal, i.e., $S(\rho_ {\rm color})=S (\rho_ {\rm position})$. If taking the partial trace operation with respect to the color qubit $q_{\rm c}$, we are able to obtain the following density matrix $\rho _{\rm position}$
\begin{equation}
\begin{array}{l}
\rho _{\rm 1_{p}2_{p}}=\text{Tr}_{\rm c}(\rho_{\rm color, position} )\\
~~~~~~~=\left(
\begin{array}{cccc}
 0.250 & 0.146 & 0.146 & 0.250 \\
 0.146 & 0.250 & 0.250 & 0.146 \\
 0.146 & 0.250 & 0.250 & 0.146 \\
 0.250 & 0.146 & 0.146 & 0.250 \\
\end{array}
\right).
\end{array}
\end{equation}
This reduced matrix is no longer a pure state since $\text {Tr} (\rho_{1_{p}2_{p}} ^ 2) \neq 1$. The von Neumann entropy is equal to
$S(\rho _{1_{p}2_{p}})=-\text{Tr}(\rho _{1_{p}2_{p}} \log \rho _{1_{p}2_{p}})=0.509$.
A detailed comparison of the entanglement between the qubit of color and the qubits of position for binary images of size $2\times 2$ is given below (see Table 1):

\begin{table}[H]
\begin{center}
\begin{tabular}{|c|c|c|c|c|c|c|c|c|c|}
\hline
\rm {Image}&$\text{Tr}(\rho_{\rm cp}^{2})$ & $\text{Tr}(\rho _{\rm c}^2)$ & $\text{Tr}(\rho _{\rm p}^2)$ & $S(\rho_{\rm c})$ & $S(\rho_{\rm p})$  \\ \hline
\includegraphics[width=1.5cm]{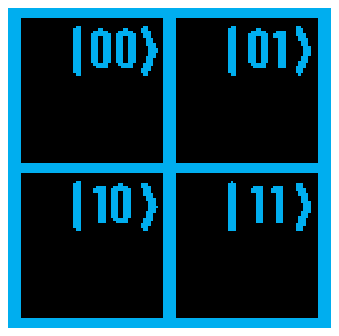}&1 & 1 & 1 & 0 & 0 \\ \hline
\includegraphics[width=1.5cm]{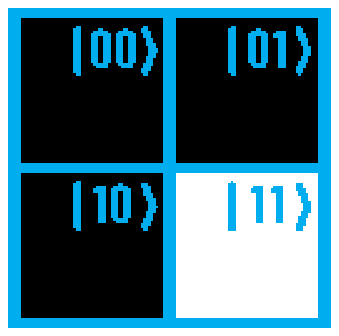} &1 & 0.625 & 0.625 & 0.811 & 0.811\\ \hline
\includegraphics[width=1.5cm]{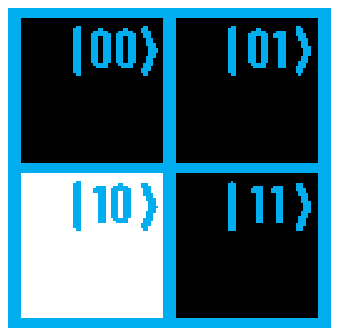}&1 & 0.625 & 0.625 & 0.811  & 0.811\\ \hline
\includegraphics[width=1.5cm]{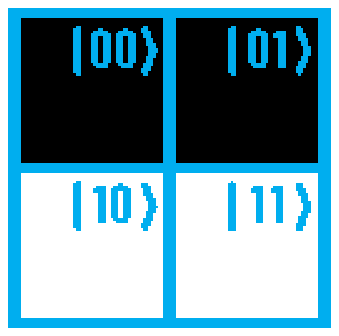}&1 & 0.500 & 0.500 & 1.000  & 1.000\\ \hline
\includegraphics[width=1.5cm]{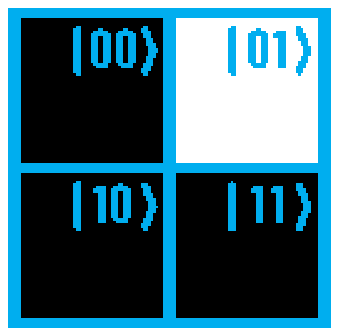}&1 & 0.625 & 0.625 & 0.811  & 0.811\\ \hline
\includegraphics[width=1.5cm]{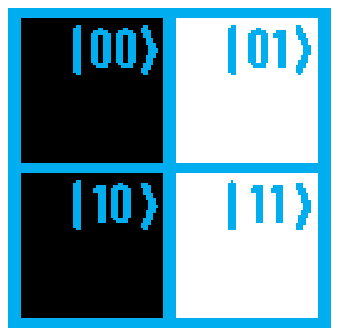}&1 & 0.500 & 0.500 & 1.000  & 1.000\\ \hline
\includegraphics[width=1.5cm]{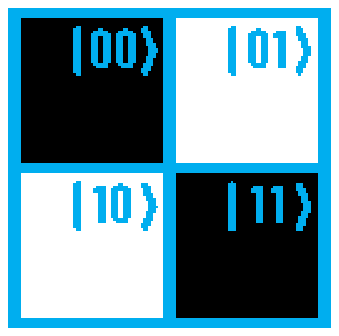}&1 & 0.500 & 0.500 & 1.000  & 1.000\\ \hline
\end{tabular}
\end{center}
\label{tab:entrindividual}
\end{table}

\begin{table}[H]
\begin{center}
\begin{tabular}{|c|c|c|c|c|c|c|c|c|c|}
\hline
\includegraphics[width=1.5cm]{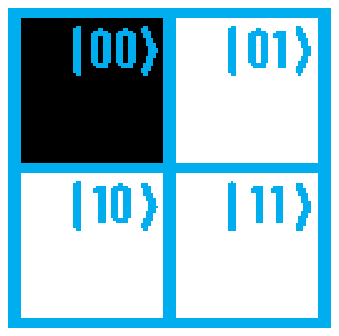}&1 & 0.625 & 0.625 & 0.811  & 0.811\\ \hline
\includegraphics[width=1.5cm]{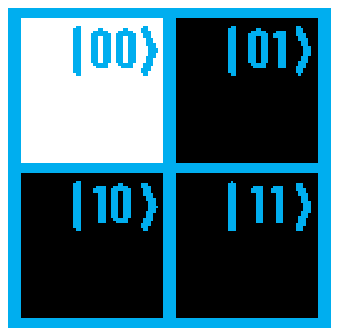}&1 & 0.625 & 0.625 & 0.811  & 0.811\\ \hline
\includegraphics[width=1.5cm]{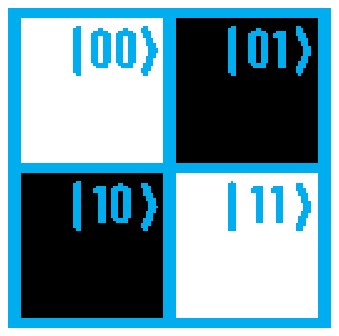}&1 & 0.500 & 0.500 & 1.000  & 1.000\\ \hline
\includegraphics[width=1.5cm]{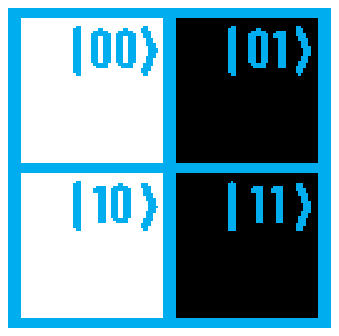}&1 & 0.500 & 0.500 & 1.000  & 1.000\\ \hline
\includegraphics[width=1.5cm]{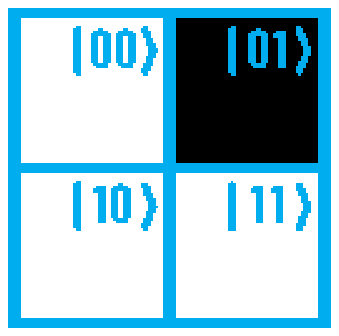}&1 & 0.625 & 0.625 & 0.811  & 0.811\\ \hline
\includegraphics[width=1.5cm]{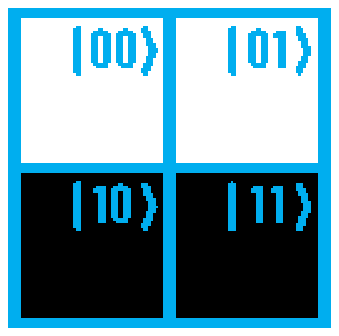}&1 & 0.500 & 0.500 & 1.000  & 1.000\\ \hline
\includegraphics[width=1.5cm]{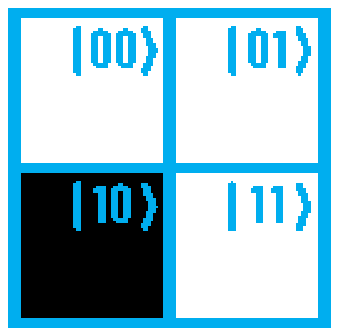}&1 & 0.625 & 0.625 & 0.811  & 0.811\\ \hline
\includegraphics[width=1.5cm]{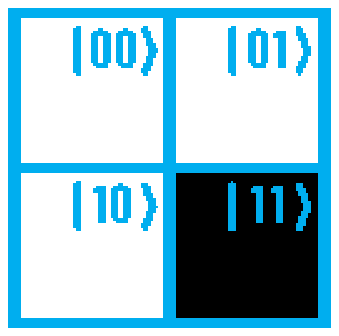}&1 & 0.625 & 0.625 & 0.811  & 0.811\\ \hline
\includegraphics[width=1.5cm]{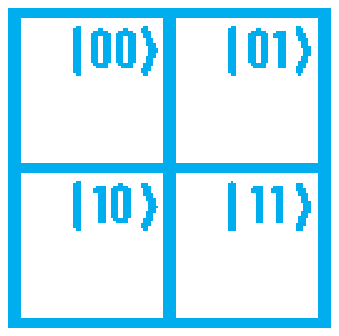}&1 & 1 & 1 & 0  & 0 \\ \hline
\end{tabular}
\end{center}
\caption{Entanglement measures for a 2D binary image of 4 pixels.}
\end{table}

\section{Process of comparing two images}
\begin{figure}[h]
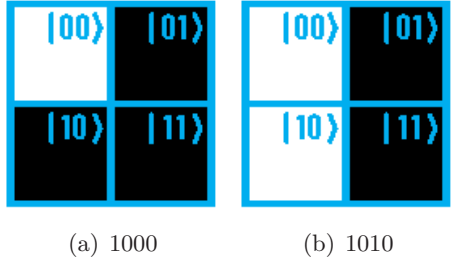

\centering
\subfigure[1000]{\includegraphics[width=3cm]{Im1000.eps}}
\subfigure[1010]{\includegraphics[width=3cm]{Im1010.eps}}
\caption{{Comparison of two images by means of mutual information measures.}}  \label{fig:dosimagenes}
\end{figure}

Now, we are going to compare two images. The wave function is constructed by the color information of two images, say $A$ and $B$, as well as the corresponding position information. We will calculate the individual and joint entropies for the subsystem and the mutual information by choosing two images as shown in Fig. \ref{fig:dosimagenes}.

Before studying them, let us define the quantum mutual information as
\begin{equation}\label{iccuantica}
I(A;B)=S(A)+S(B)-S(A, B)=S(\rho_{AB}||\rho_{A}\otimes \rho_{B}),
\end{equation}
where $S(\rho||\gamma)=\text{Tr}(\rho \log \rho-\rho \log \gamma)$. This measure is non-negative and will be zero only if the product state $\rho_{AB}=\rho_{A} \otimes \rho_{B}$.

For a bipartite pure state $\rho_{AB}$,
$S(A|B)=-S(A)=-S(B) < 0$. When the systems $ A, B$ are perfectly correlated in classical case, then $ H (A, B) = H (A) = H (B) $, and the conditional entropy vanishes, $ H (A | B) = H (B | A) = 0 $. The mutual information parameter $ I (X; Y) = H (X) = H (Y) $ in the classic case\footnote{We often use $S$ and $H$ denote the quantum and classical entropy respectively.}. This means that the mutual information for a classical system with two correlated systems is equal to the entropy of one of the subsystems. For a pure bipartite state $AB$, however, the mutual information leads to an important relation
\begin{equation}
I(A;B)=2S(A)= 2S(B),
\end{equation}which implies that the quantum mutual information is double value of the quantum entropy. Therefore, the quantum correlations are stronger than the classical correlations.

On the other hand, some important identities in tripartite about joint entropy and mutual information are expressed as \cite{Lau, sazim_quantum_2016, phoenix_quantum_2015}
\begin{equation}\label{eq:interaccion}
I_{0}(A;B;C)=I(A;B)-I(A;B|C),
\end{equation}
where $I(A;B|C)=S(A|C)+S(B|C)-S(A, B|C)$;
\begin{equation}\label{eq:correlaciontotal}
I_{T}(A;B;C)=I(A;B)+I(AB;C),
\end{equation}
where $I(AB;C)=I(A;C)+I(B;C|A)$ and
\begin{equation}\label{eq:correlaciondual}
I_{D}(A;B;C)=I(A;BC)+I(B;C|A),
\end{equation}
where $I(A;BC)=I(A;B)+I(A;C|B)$. They represent the interaction information $I_{0}$, total correlation $I_{T}$ and total dual correlation $I_{D}$ respectively in the tripartite system.  As far as the conditional entropy, we might generalize the classical form $H(A|B)=H(A, B)-H(B) \geq 0$ to quantum case $S(A|B)=S(A, B)-S(B)$, which could be negative for quantum states.

Now, we calculate the individual and joint entropies for the subsystem and the mutual information for the images shown in Figure \ref{fig:dosimagenes}. In this case, we have the color vectors:
\begin{equation*}
\vec\theta_{A}= \begin{bmatrix}
 1.571 \\0\\  0\\  0
\end{bmatrix}
\mbox {     ,     }
\vec\theta_{B}= \begin{bmatrix}
 1.571 \\0\\  1.571\\  0
\end{bmatrix}
\end{equation*}
with the corresponding color qubits
\begin{equation*}
q_{A}= \begin{bmatrix}
 \left|1_{A}\right\rangle \\\left|0_{A}\right\rangle \\  \left|0_{A}\right\rangle \\  \left|0_{A}\right\rangle
\end{bmatrix}
\mbox {     ,     }
q_{B}= \begin{bmatrix}
 \left|1_{B}\right\rangle \\\left|0_{B}\right\rangle \\  \left|1_{B}\right\rangle \\  \left|0_{B}\right\rangle
\end{bmatrix}.
\end{equation*}

The normalized wave function can be constructed as
\begin{equation}
\begin{array}{l}
|\psi\rangle =\displaystyle\frac{1}{2} \sum\limits_{i=0}^{3} |i \rangle \otimes q_{A}\otimes q_{B}\\[3mm]
~~~~~~~ = 0.5 \left|001_{A}1_{B}\right\rangle +0.5 \left|010_{A}0_{B}\right\rangle +0.5 \left|100_{A}1_{B}\right\rangle +0.5 \left|110_{A}0_{B}\right\rangle,
\end{array}
\end{equation} from which we have the density matrix
\begin{equation}
\rho_{12AB} =\frac{1}{4}\left(
\begin{array}{cccccccccccccccc}
 0 & 0 & 0 & 0 & 0 & 0 & 0 & 0 & 0 & 0 & 0 & 0 & 0 & 0 & 0 & 0 \\
 0 & 0 & 0 & 0 & 0 & 0 & 0 & 0 & 0 & 0 & 0 & 0 & 0 & 0 & 0 & 0 \\
 0 & 0 & 0 & 0 & 0 & 0 & 0 & 0 & 0 & 0 & 0 & 0 & 0 & 0 & 0 & 0 \\
 0 & 0 & 0 & 1 & 1 & 0 & 0 & 0 & 0 & 1 & 0 & 0 & 1 & 0 & 0 & 0 \\
 0 & 0 & 0 & 1 & 1 & 0 & 0 & 0 & 0 & 1 & 0 & 0 & 1 & 0 & 0 & 0. \\
 0 & 0 & 0 & 0 & 0 & 0 & 0 & 0 & 0 & 0 & 0 & 0 & 0 & 0 & 0 & 0 \\
 0 & 0 & 0 & 0 & 0 & 0 & 0 & 0 & 0 & 0 & 0 & 0 & 0 & 0 & 0 & 0 \\
 0 & 0 & 0 & 0 & 0 & 0 & 0 & 0 & 0 & 0 & 0 & 0 & 0 & 0 & 0 & 0 \\
 0 & 0 & 0 & 0 & 0 & 0 & 0 & 0 & 0 & 0 & 0 & 0 & 0 & 0 & 0 & 0 \\
 0 & 0 & 0 & 1 & 1 & 0 & 0 & 0 & 0 & 1 & 0 & 0 & 1 & 0 & 0 & 0 \\
 0 & 0 & 0 & 0 & 0 & 0 & 0 & 0 & 0 & 0 & 0 & 0 & 0 & 0 & 0 & 0 \\
 0 & 0 & 0 & 0 & 0 & 0 & 0 & 0 & 0 & 0 & 0 & 0 & 0 & 0 & 0 & 0 \\
 0 & 0 & 0 & 1 & 1 & 0 & 0 & 0 & 0 & 1 & 0 & 0 & 1 & 0 & 0 & 0 \\
  0 & 0 & 0 & 0 & 0 & 0 & 0 & 0 & 0 & 0 & 0 & 0 & 0 & 0 & 0 & 0 \\
  0 & 0 & 0 & 0 & 0 & 0 & 0 & 0 & 0 & 0 & 0 & 0 & 0 & 0 & 0 & 0 \\
 0 & 0 & 0 & 0 & 0 & 0 & 0 & 0 & 0 & 0 & 0 & 0 & 0 & 0 & 0 & 0 \\
\end{array}
\right).
\end{equation}
According to this density, we are able to calculate all relevant von Neumann entropies of the subsystems\footnote{It should be emphasized that the entropy $S$ is calculated by considering the  binary logarithm (the base is taken $2$ but not natural number $e$), and the units of entropy $S$ are expressed in bits.}.
\begin{itemize}
\item $\rho _{12A}=\text{Tr}_{B}(\rho_{12AB} )$
\begin{equation}
\rho _{12A}=\left(
\begin{array}{cccccccc}
 0 & 0 & 0 & 0 & 0 & 0 & 0 & 0 \\
 0 & 1 & 0 & 0 & 1 & 0 & 0 & 0 \\
 0 & 0 & 1 & 0 & 0 & 0 & 1 & 0 \\
 0 & 0 & 0 & 0 & 0 & 0 & 0 & 0 \\
 0 & 1 & 0 & 0 & 1 & 0 & 0 & 0 \\
 0 & 0 & 0 & 0 & 0 & 0 & 0 & 0 \\
 0 & 0 & 1 & 0 & 0 & 0 & 1 & 0 \\
 0 & 0 & 0 & 0 & 0 & 0 & 0 & 0 \\
\end{array}
\right), ~~~~S(\rho _{12A})=1.0.
\end{equation}
\end{itemize}

\begin{itemize}
\item $\rho_{12B}=\text{Tr}_{A}(\rho_{12AB} )$
\begin{equation}
\rho _{12B}=\frac{1}{4}\left(
\begin{array}{cccccccc}
 0 & 0 & 0 & 0 & 0 & 0 & 0 & 0 \\
 0 & 1 & 0 & 0 & 0 & 0 & 0 & 0 \\
 0 & 0 & 1 & 0 & 0 & 1 & 1 & 0 \\
 0 & 0 & 0 & 0 & 0 & 0 & 0 & 0 \\
 0 & 0 & 0 & 0 & 0 & 0 & 0 & 0 \\
 0 & 0 & 1 & 0 & 0 & 1 & 1 & 0 \\
 0 & 0 & 1 & 0 & 0 & 1 & 1 & 0 \\
 0 & 0 & 0 & 0 & 0 & 0 & 0 & 0 \\
\end{array}
\right), ~~~S(\rho_{12B})=0.811.
\end{equation}
\item $\rho_{12}=\text{Tr}_{A}\left(\text{Tr}_{B}(\rho_{12AB} )\right)$
\begin{equation}
\rho_{12}=\frac{1}{4}\left(
\begin{array}{cccc}
 1 & 0 & 0 & 0 \\
 0 & 1 & 0 & 1 \\
 0 & 0 & 1 & 0 \\
 0 & 1 & 0 & 1 \\
\end{array}
\right), ~~~S(\rho_{12})=1.5.
\end{equation}
\item$\rho _{AB}=\text{Tr}_{2}\left(\text{Tr}_{1}(\rho_{12AB} )\right)$
\begin{equation}
\rho _{AB}=\left(
\begin{array}{cccc}
 0.5 & 0 & 0 & 0 \\
 0 & 0.25 & 0 & 0 \\
 0 & 0 & 0 & 0 \\
 0 & 0 & 0 & 0.25 \\
\end{array}
\right), ~~~~S(\rho _{AB})=1.5.
\end{equation}
\item $\rho_{A}=\text{Tr}_{B}\left(\rho _{AB}\right)$
\begin{equation}
\rho_{A}=\left(
\begin{array}{cc}
 0.75 & 0 \\
 0 & 0.25 \\
\end{array}
\right), ~~~~S(\rho_{A})=0.811.
\end{equation}
\item $\rho_{B}=\text{Tr}_{A}\left(\rho _{AB}\right)$
\begin{equation}
\rho_{B}=\left(
\begin{array}{cc}
 0.5 & 0 \\
 0 & 0.5 \\
\end{array}
\right), ~~~S(\rho_{B})=1.0.
\end{equation}
\end{itemize}

We are now in the position to compare the images of size $2\times 2$ with the patron configuration image $\{0, 255, 255, 255\}$. The corresponding results are given in the following Table 2.

\begin{table}[H]
\begin{center}
{\small
\begin{tabular}{|c|c|c|c|c|c|c|c|c|c|}
\hline
{\rm image} & $S(A)$ & $S(B)$ & \multicolumn{1}{l|}{$S(12)$} & \multicolumn{1}{l|}{$S(A, B)$} & \multicolumn{1}{l|}{$S(A, 12)$} & \multicolumn{1}{l|}{$S(B, 12)$} & \multicolumn{1}{l|}{$I_{0}$} & \multicolumn{1}{l|}{$I_{T}$} & \multicolumn{1}{l|}{$I_{D}$} \\ \hline \hline
\includegraphics[width=.15\textwidth]{Im0000.eps} & 0.811 & 0.000 &  0.811 &  \cellcolor[rgb]{0., 0.9, 0.9} 0.811 & 0 & 0.811 & 0 & 1.622 & 1.622 \\ \hline
\includegraphics[width=.15\textwidth]{Im0001.eps} & 0.811 & 0.811 & 1.5 & 1.5 & 0.811 & 0.811 & 0 & 3.122 & 3.122 \\ \hline
\includegraphics[width=.15\textwidth]{Im0010.eps} & 0.811 & 0.811 & 1.5 & 1.5 & 0.811 & 0.811 & 0 & 3.122 & 3.122\\ \hline
\includegraphics[width=.15\textwidth]{Im0011.eps} & 0.811 & 1.000 & 1.5 & 1.5 & 1.000 & 0.811 & 0 & 3.311 & 3.311 \\ \hline
\includegraphics[width=.15\textwidth]{Im0100.eps} & 0.811 & 0.811 & 1.5 & 1.5 & 0.811 & 0.811 & 0 & 3.122 & 3.122 \\ \hline
\includegraphics[width=.15\textwidth]{Im0101.eps} & 0.811 & 1.000 & 1.5 & 1.5 & 1.000 & 0.811 & 0 & 3.311 & 3.311 \\ \hline
\includegraphics[width=.15\textwidth]{Im0110.eps} & 0.811 & 1.000 & 1.5 & 1.5 & 1.000 & 0.811 & 0 & 3.311 & 3.311 \\ \hline
\end{tabular}
}
\end{center}
\label{tab:EntConjunta1000}
\end{table}

\begin{table}[H]
\begin{center}
{\small
\begin{tabular}{|c|c|c|c|c|c|c|c|c|c|}
\hline
\includegraphics[width=.15\textwidth]{Im0111.eps} & 0.811 & 0.811 & 0.811 & \cellcolor[rgb]{0., 0.9, 0.9}0.811 & 0.811 & 0.811 & 0 & \cellcolor[rgb]{0., 0.8, 0.2}2.433 & \cellcolor[rgb]{0., 0.8, 0.2}2.433 \\ \hline
\includegraphics[width=.15\textwidth]{Im1000.eps} & 0.811 & 0.811 & 0.811 & \cellcolor[rgb]{0., 0.9, 0.9}0.811 & 0.811 & 0.811 & 0 & \cellcolor[rgb]{0., 0.8, 0.2}2.433 & \cellcolor[rgb]{0., 0.8, 0.2}2.433 \\ \hline
\includegraphics[width=.15\textwidth]{Im1001.eps} & 0.811 & 1.000 & 1.5 & 1.5 & 1.000 & 0.811 & 0 & 3.311 & 3.311 \\ \hline
\includegraphics[width=.15\textwidth]{Im1010.eps} & 0.811 & 1.000 & 1.5 & 1.5 & 1.000 & 0.811 & 0 & 3.311 & 3.311 \\ \hline
\includegraphics[width=.15\textwidth]{Im1011.eps} & 0.811 & 0.811 & 1.5 & 1.5 & 0.811 & 0.811 & 0 & 3.122 & 3.122 \\ \hline
\includegraphics[width=.15\textwidth]{Im1100.eps} & 0.811 & 1.000 & 1.5 & 1.5 & 1.000 & 0.811 & 0 & 3.311 & 3.311 \\ \hline
\includegraphics[width=.15\textwidth]{Im1101.eps} & 0.811 & 0.811 & 1.5 & 1.5 & 0.811 & 0.811 & 0 & 3.122 & 3.122 \\ \hline
\includegraphics[width=.15\textwidth]{Im1110.eps} & 0.811 & 0.811 & 1.5 & 1.5 & 0.811 & 0.811 & 0 & 3.122 & 3.122 \\ \hline
\includegraphics[width=.15\textwidth]{Im1111.eps}  & 0.811 & 0.000 &  0.811 &  \cellcolor[rgb]{0., 0.9, 0.9} 0.811 & 0 & 0.811 & 0 & 1.622 & 1.622 \\ \hline
\end{tabular}
}
\caption{We make an exhaustive comparison between two binary images. We use an image of 4 pixels in the  configuration $\{0, 255, 255, 255\}$ and the 16 possible configurations to compare. We calculate the individual and dual entropies and the information interaction $I_{0}$, the total correlation $I_{T}$ and the total dual correlation $I_{D}$. We can see that for minimal measures of the joint entropy between the qubits that encode the color information of the images $A, B$ (in blue), there exist maximal amounts of $I_{T}$, $I_{D}$ (in green) that point to the optimal register.}
\end{center}
\label{tab:EntConjunta1000}
\end{table}
On the other hand, based on Eqs. (\ref{eq:interaccion}), (\ref{eq:correlaciontotal}) and (\ref{eq:correlaciondual}) we are able to calculate $I_{0}$, $I_{T}$ and $I_{D}$ for the present system $\rho_{AB12}$ as
$I_{0}(A; B; 12)=S(A)+S(B)+S(12)-S(A, B)-S(A, 12)-S(B, 12)+S(A, B, 12)=0$, $I_{T}= S(A)+S(B)+S(12)-S(A, B, 12)=3.311$ and $I_{D}=S(A, B)+S(A, 12)+S(B, 12)-2S(A, B, 12)=3.311$.

We might expect to show several significant differences between bipartite and tripartite correlations. First, we note that the information interaction $I_{0}$ is always zero regardless of the configuration since the complete system we are studying is a pure state. Second, we notice that the total correlation $I_{T}$ and the total dual correlation $I_{D}$ are equal to each other.

\section{Comparison of two images of $2\times 2$ pixels under change of color of one pixel and of $8 \times 8$ pixels under translation transform based on classical and quantum methods}

The color of the pixels affects the quantum entropy of the joint system. This is an important difference with respect to the classical method which is invariant under color pixel transformations.

\begin{figure}[H]
\begin{center}
\includegraphics[width=12cm]{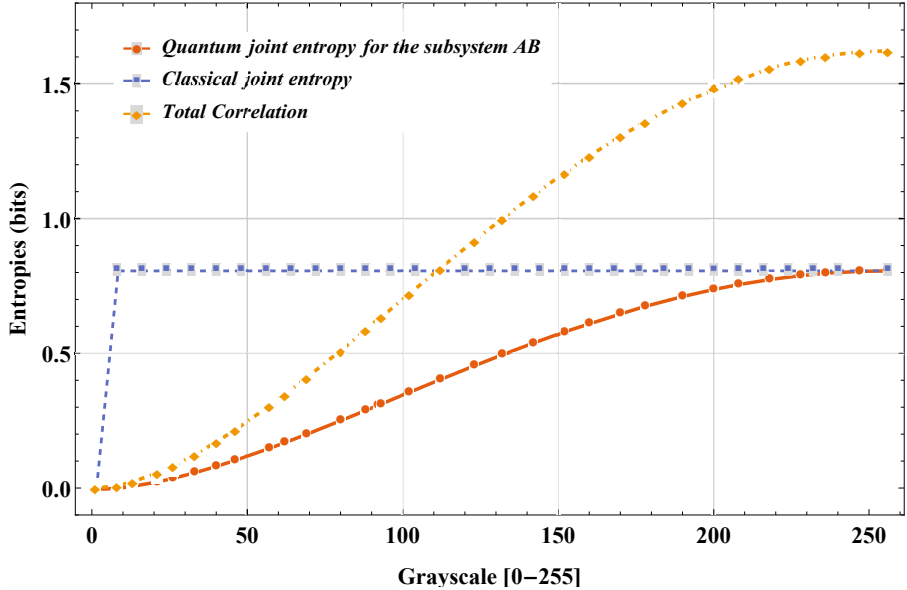}
\caption{{Comparison between the quantum entropy of the subsystem AB, the quantum total interaction ($I_T$), and the classical joint entropy of the images A in the configuration $\{0, 0, 0, 0\}$ and B in the configuration $\{0, 0, X, 0\}$, when we change the value of the pixel $X$ of the image B from [0-255].}\label{fig:1}}
\end{center}
\end{figure}

\begin{figure}[H]
\begin{center}
\includegraphics[width=12cm]{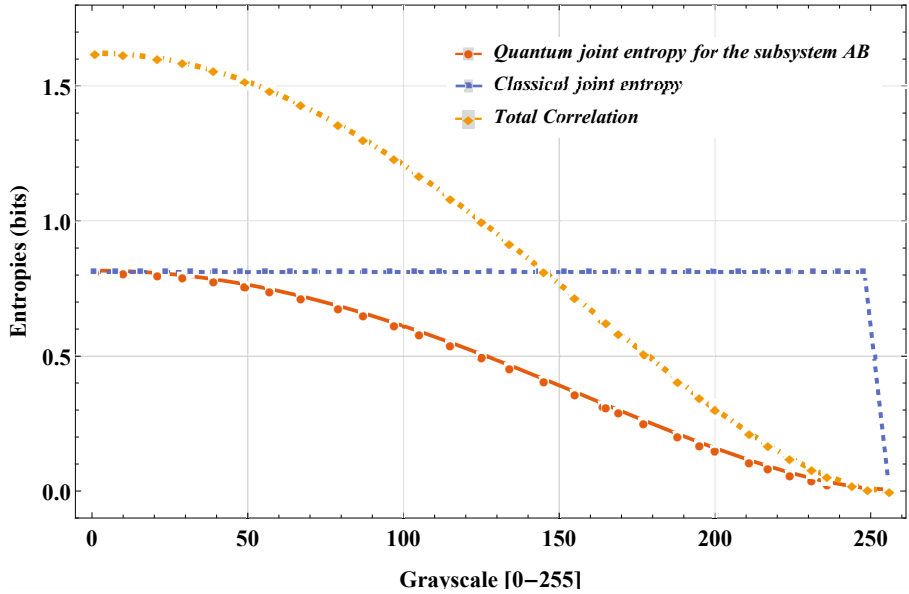}
\caption{{Comparison between the quantum entropy of the subsystem AB, the quantum total interaction ($I_T$), and the classical joint entropy of the images A in the configuration $\{255, 255, 255, 255\}$ and B in the configuration $\{255, 255, X, 255\}$, when we change the value of the pixel $X$ of the image B from [0-255] }\label{fig:2}}
\end{center}
\end{figure}

We can see in Figure \ref{fig:1} that the amount of quantum joint entropy for the subsystem $AB$ can arrive to the value of the classical joint entropy, but the total correlation $I_T$ can reach the double amount of the quantum joint entropy. In the symmetric case by changing the initial configuration for example with all the pixels with color index of $255$ and changing the color of one pixel from [0-255] we obtain the entropy as shown in Figure \ref{fig:2}

\begin{figure}[H]
\begin{center}
\includegraphics[width=12cm]{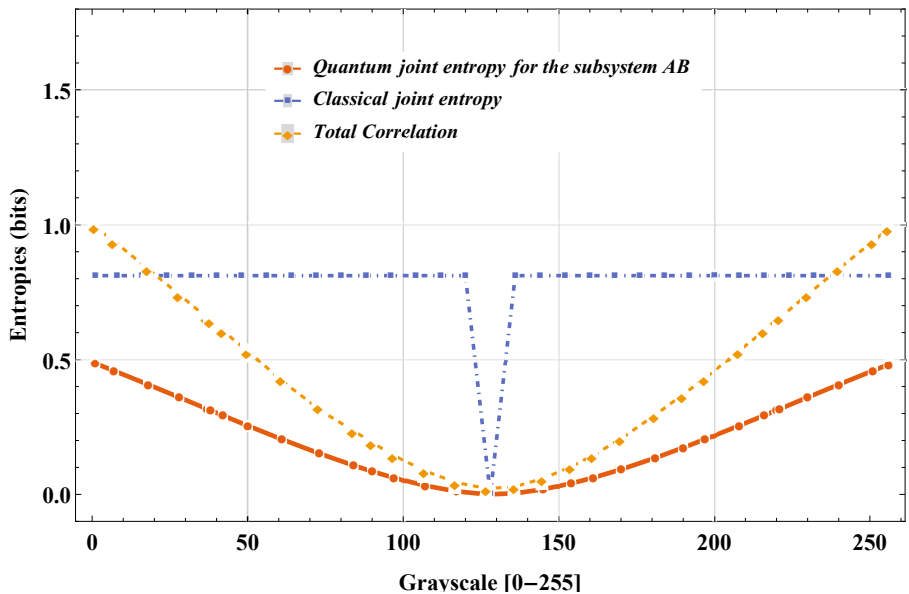}
\caption{{Comparison between the quantum entropy of the subsystem AB, the quantum total interaction ($I_T$), and the classical joint entropy of the images A in the configuration $\{128, 128, 128, 128\}$ and B in the configuration $\{128, 128, X, 128\}$, when we change the value of the pixel X of the image B from [0-255]. }\label{fig:3}}
\end{center}
\end{figure}

For an image configuration where all pixels for both images have a color index of $128$, and we change the color of one pixel of the image B from [0-255], we have the relations as illustrated in Figure \ref{fig:3}

We can see that the classical joint entropy is invariant under transformation of change of color. This is because the joint histogram contains the information about the number of correspondences between pixels of both images. When there exists a perfect correlation between the images and all the pixels which are in color index 128, we find that their entropy is zero as a minimum value. However, the quantum measures are sensitive under transformation of change of color. In the previous Figure \ref{fig:3} we only plot the quantum entropy of the subsystem $\rho_{AB}$, without the qubits that code the position, and the mutual information entropy for the tripartite system $I_T$. From this Fig.6  we can see that the amount of quantum joint entropy for the subsystem $AB$ does not arrive to the value of the classical joint entropy. The total correlation $I_T$  can also reach the double amount of the classical joint entropy.

\begin{figure}[H]
\begin{center}
\includegraphics[width=5cm]{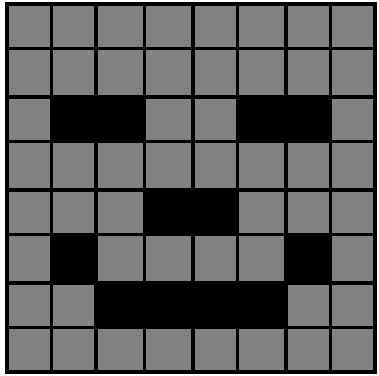}
\caption{{Patron image to compare}\label{fig:4}}
\end{center}
\end{figure}

When we extend the size of the image to $8\times8$ pixels, we can see more details about the differences between the classical parameters and the quantum parameters. We compare a patron image (see Fig.\ref{fig:4}) with the same image but translated one pixel for iteration.

\begin{figure}[H]
\begin{sideways}{90}
\includegraphics[width=20cm]{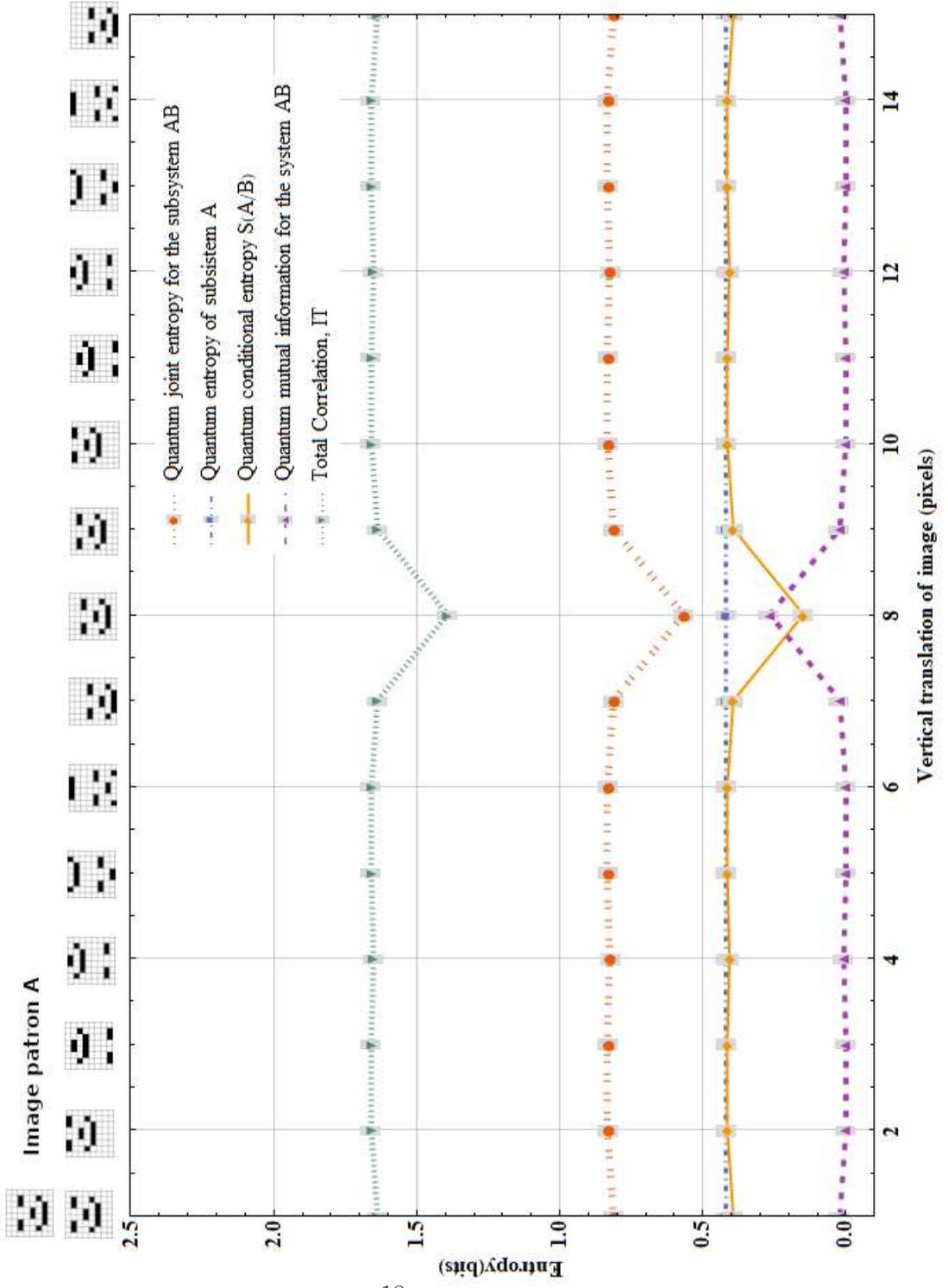}
\end{sideways}
\caption{{Quantum entropies calculated for a comparison of images of size $8\time 8$. The gray color (128) is used instead of white color.  }\label{5}}
\end{figure}

\begin{figure}[H]
\begin{sideways}{90}
\includegraphics[width=20cm]{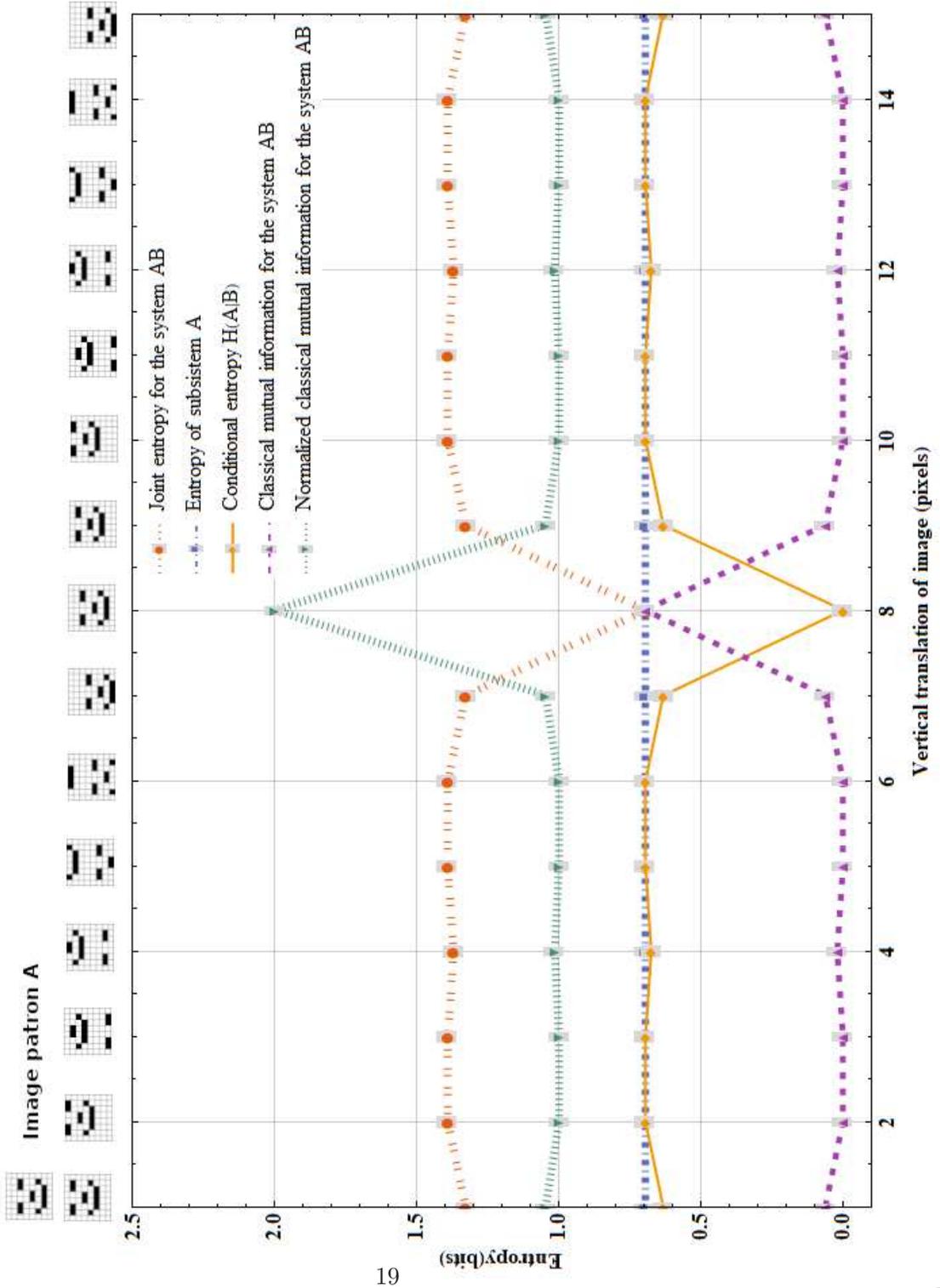}
\end{sideways}
\caption{{Classical entropies calculated for a comparison of images of size $8\time 8$. The gray color (128) is used instead white color.  }\label{6}}
\end{figure}

If we use a binary configuration with only white and black colors, the quantum and classical measures are the same, except for the total correlation $I_T$, which is a quantum measure affected by the condition whether it is calculated in pure states or not. When we calculate the classical and quantum measures for a patron image configuration with gray colors (128) instead of white color, we obtain the next two graphics \ref{5} and \ref{6}.

First, we note that the classical measures are like the joint entropy, the mutual information are more sensitive to the optimal register than that of the quantum measures. The green line is the normalized classical mutual information. In contrast, the quantum measures are less sensitive to the optimal register. The quantum mutual information of the system $\rho_{AB}$ has a minimum value in the optimal register. Also, we can see that the Total Correlation ($I_T$) has more entropy than the classical mutual information.

\section{Conclusions}

The production and manipulation of correlated systems of qubits where the quantum nature of the correlation can be used as a resource to yield properties unavailable within a classical framework is a very active and important area of research. It would seem therefore that understanding the nature of the correlation between quantum systems is an important goal.
In the case the concept of a joint histogram is very useful since it offers a mathematical structure where we can observe the correlations between random variables that may correspond to very different aspects of the phenomenon studied. The quantum form in which the problem is coded offers a new dimension that is sensitive to the details that the joint histogram is not sensible. In this work, the results have been compared between two images of size $2\times 2$ and $8\times 8$ from different classical and quantum methods. We find that the classical joint entropy is invariant under transformation of change of color but the quantum entropy is sensitive to this change. It is shown that the total correlation $I_T$ could arrive to the double amount of the classical joint entropy. Before ending this work, we would like to give a useful remark on the configurations of the images. As we know, by exchanging the black $0$ and white $1$, the entanglement measures such as the entropy is same. For example, the images 1001 and 0110 have same characteristics as shown in Table 2. We are going to explore other measures to distinguish them in the near future.

\vskip 5mm
\noindent
{\Large \bf Competing Interests}\\
The authors declare that there is no conflict of interests regarding the publication of this paper.

\vskip 5mm
\noindent
{\Large \bf Data Availability }: No data was used to support this study.

\section*{Acknowledgments} This work is supported by project 20180677-SIP-IPN, COFAA-IPN, Mexico and partially by the CONACYT project under grant
No. 288856-CB-2016.

\bibliographystyle{unsrt}

\end{document}